\documentstyle[prl,aps,twocolumn,psfig]{revtex}

\begin{document}
\twocolumn[\hsize\textwidth\columnwidth\hsize\csname
@twocolumnfalse\endcsname

\draft

\title{Limitations to bit-rate and spatial capacity of an optical data transmission channel}
\author{Er'el Granot\footnotemark and Shmuel Sternklar}

\address{Department of Electrical Engineering, Academic College  of Judea and Samaria, Ariel 3, Israel}

\date{\today}
\maketitle
\begin{abstract}
\begin{quote}
\parbox{16 cm}{\small
The maximum bit-rate of a slab waveguide is ultimately determined
by the waveguide dispersion. We show that while the maximum bit
rate in a waveguide is inversely proportional to the waveguide's
width, bit rate per unit width (i.e., spatial capacity)
\emph{decreases}, and in the limit of a zero-width waveguide it
converges to $\sim \omega/\sqrt(\lambda L)$ (where $L$ is the
length and $\omega$ and $\lambda$ are the beam's frequency and
wavelength respectively). This result, which is independent of the
waveguide's refractive indices, is qualitatively equivalent to the
transmission rate per unit of width in free space. We also show
that in a 3D waveguide (e.g., fibers), unlike free space, the
spatial capacity vanishes in the same limit.}
\end{quote}
\end{abstract}


]

\narrowtext \footnotetext{erel.g@kailight.com} \noindent
The importance of maximizing the transmission rate of
communication systems is well recognized. There is also a growing
interest in short-distance, yet high bit-rate, wireless systems
\cite{leeper}. However, due to the beam divergence in free space
the bit-rate is not the only parameter relevant to data
transmission. The amount of data concentration in space, known as
spatial capacity(SC), is also a crucial parameter.

There are many effects which impair information transfer in
optical waveguides (slab waveguides or optical fibers), such as
noise, absorption, scattering, dispersion and nonlinear effects.
According to the Shannon formula \cite{Shannon} and its more
applied derivatives (see, for example
\cite{Tang,Neifeld,Narimanov_Mitra}) an information channel's
maximum bit rate is mostly determined by the channel's noise.

In the presence of Gaussian noise, the main problem encountered in
high data rate transmission is dispersion. In practice the highest
bit-rate is determined by the criterion that the dispersion
broadened pulse (which represents a digital bit) should not exceed
its allocated slot\cite{Agrawal}.

In a single-mode waveguide the dispersion is caused by mainly
three different factors: material dispersion (MD), polarization
dispersion (PD) and waveguide dispersion (WD). In the case of a
wide waveguide, the WD is negligible, and the MD is the dominant
factor (usually the influence of PD is smaller) and absolutely
large. However, if the cladding of the waveguide is made of free
space then the influence of the MD (and of PD) will decrease by
narrowing the waveguide. It so happens, however, that reducing the
waveguide width also decreases the WD (even though it is still the
dominant factor). Therefore, by reducing the waveguide's width,
higher bit rates are possible.

It seems, therefore, that the best transmission rate will be
achieved for a waveguide with zero width. This is a strange
consequence since a zero-width waveguide should not behave
qualitatively different than free space.

In this paper we will show that while the transmission rate of a
single slab waveguide does increase when its width shrinks, the SC
decreases, and in fact the SC for zero-width waveguides is
qualitatively similar to the SC of free space. However, in three
dimensions (e.g., fiber) this is \emph{not} the case and a 2D
array of zero-width fibers has a different SC than that of free
space.

We begin with the simple model of a slab waveguide. The index of
refraction of such a waveguide can be written 

\begin{equation}
n(x)=\left\{\begin{array}{cc}
  n_1 & |x|<a \\
  n_2 & |x|>a
\end{array} \right.
\end{equation}
where $2a$ is the waveguide's width.

For simplicity we choose the TE mode to describe wave propagation
in the waveguide, i.e.,
\begin{equation}
{\mathbf{E}}(x,z)=\hat{y}\psi(x,z)\exp(-i\omega t) \label{E_field}
\end{equation}

where $\psi(x,z)$ satisfies the wave equation

\begin{equation}
\nabla^2\psi+k^2\psi=0 \label{wave_eq}
\end{equation}

and $k$ (wave number), $\omega$ (angular frequency) and $\lambda$
(wavelength) are related according to

\begin{equation}
k=\frac{n\omega}{c}=\frac{2\pi n}{\lambda}.
\end{equation}


If we use $\beta$ to represent the wave propagation constant along
the waveguide, then

\begin{equation}
\psi(x,z)=\varphi(x)\exp(i\beta z)
\end{equation}

with the simple 1D equation for $\varphi$

\begin{equation}
\frac{d^2\varphi(x)}{dx^2}+\left(k^2-\beta^2\right)\varphi(x)=0.
\end{equation}

The stationary solution is equivalent to a simple eigenvalue
problem. Inside the waveguide the transversal solution is
oscillatory, $~\cos \left(\sqrt{k_1^2-\beta^2} x \right)$, while
in the cladding the solution decays exponentially as $~\exp\left(-
\sqrt{\beta^2-k_2^2} x \right)$, where $k_{1,2}=2\pi
n_{1,2}/\lambda$. Matching the solution at the boundary $x=a$ we
obtain
\begin{equation}
\tan \left( a\sqrt{k_1^2-\beta^2}
\right)=\sqrt{\frac{\beta^2-k_2^2}{k_1^2-\beta^2}}. \label{eigen}
\end{equation}

We are interested in the limit of a narrow waveguide (where the
dispersion is minimal). In this regime

\begin{equation}
\beta^2-k_2^2=a^2(k_1^2-k_2^2)^2 \label{limit1}
\end{equation}

or

\begin{equation}
\beta=k_2+\frac{a^2(k_1^2-k_2^2)^2}{2k_2} \label{limit2}
\end{equation}

In order to minimize chromatic dispersion we choose air (or
vacuum) as the waveguide's cladding, i.e., $n_2=1$ and
$k_2=k_0\equiv2\pi/\lambda$. Therefore, we can use $n_1=n$ and the
propagation constant is (for $a\rightarrow 0$)

\begin{equation}
\beta=k_0[1+\frac{1}{2}(ak_0)^2(n^2-1)^2] \label{prop_constant}
\end{equation}

Thus, as the waveguide shrinks, $a\rightarrow 0$, the waveguide
dispersion decreases (since $\beta \rightarrow k_0$), and a higher
bit-rate, for a given fiber length, is attainable. One might
expect that the best transmission will be achieved for a zero
width wavelength. But this is counterintuitive, since it would
suggest that the best way to transmit information is to do so in
free space.

The problem with this reasoning lies in the fact, that the decay
length ($\xi$) outside the waveguide \emph{increases} (outside the
waveguide $\varphi(|x|>a) \sim \exp(-x/\xi)$) when the waveguide
width decreases (see Fig.1)

\begin{equation}
\xi=(\beta^2-k_2^2)^{-1/2} \rightarrow
\frac{1}{a(k_1^2-k_2^2)}=\frac{1}{ak_0^2(n^2-1)}
\end{equation}

\begin{figure}
\psfig{figure=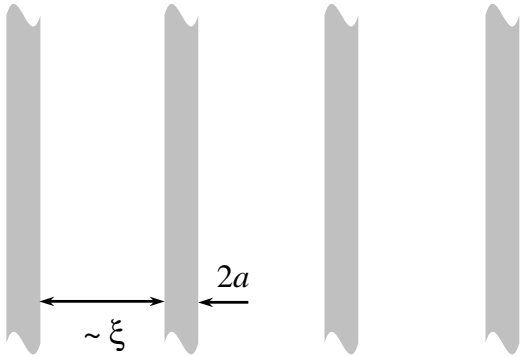,width=10cm,bbllx=100bp,bblly=200bp,bburx=650bp,bbury=550bp,clip=}
\caption{Multiple slab Waveguides}\label{fig1}
\end{figure}

and therefore, in order to avoid cross-talk between two adjacent
waveguides the distance between two such waveguides should be
\emph{increased}. This would limit the number of waveguides one
can use in a given length (or, more accurately, width).

\begin{figure}
\psfig{figure=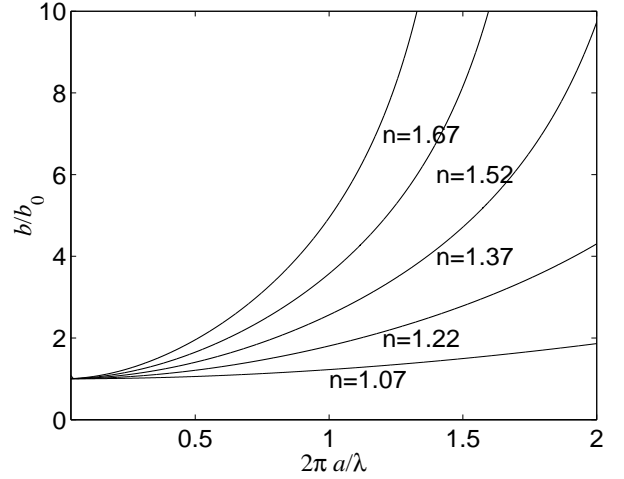,width=10cm,bbllx=1bp,bblly=-10bp,bburx=480bp,bbury=330bp,clip=}
\caption{The normalized transmission bit-rate per unit of width
($b$) for different refractive indices as a function of
$k_0a$}\label{fig2}
\end{figure}

The bit-rate ($B$) for very narrow waveguide is limited by the
waveguide dispersion \cite{Agrawal}

\begin{equation}
B<\frac{1}{4\sqrt{(d^2\beta/d\omega^2)L}}=\frac{c}{4\sqrt{(d^2\beta/dk_0^2)L}}
\end{equation}

This is the bit-rate of a single waveguide, which means that the
bit-rate per unit of width is

\begin{equation}
b=\frac{B}{\xi} \label{limit_BR}
\end{equation}

In figure 2 we plot the exact numerical solution of the
transmission bit-rate per unit of width (for different refractive
indices) as a function of the normalized width $k_0a$. Clearly,
when $a$ decreases the SC decreases, and for $k_0a \rightarrow 0$
it converges to a \emph{universal} constant:

\begin{equation}
b \rightarrow b_0 \equiv \omega \sqrt{\frac{\pi}{24\lambda L}}
\label{limit_b}
\end{equation}

This constant is universal in the sense that it is
\emph{independent} of the waveguide's characteristics (it depends
neither on its width $2a$ not on its refractive index $n$).

The limiting result (\ref{limit_b}) is consistent with the limit
of information transmission in free space, as we now show.


In free space dispersion is not a limitation, however SC will be
limited by beam divergence. For Gaussian beams, which possess the
smallest divergence, SC will be maximized when transmitting in the
confocal configuration. In this case, as shown in Fig.3, the waist
is positioned exactly at $L/2$ and is equal to $w/\sqrt{2}$, where
$w$ is the beam's width at $y=0$ and at $y=L$, $L$ is the confocal
length, and

\begin{equation}
w=2\sqrt{\frac{\lambda L}{\pi}}.
\end{equation}

\begin{figure}
\psfig{figure=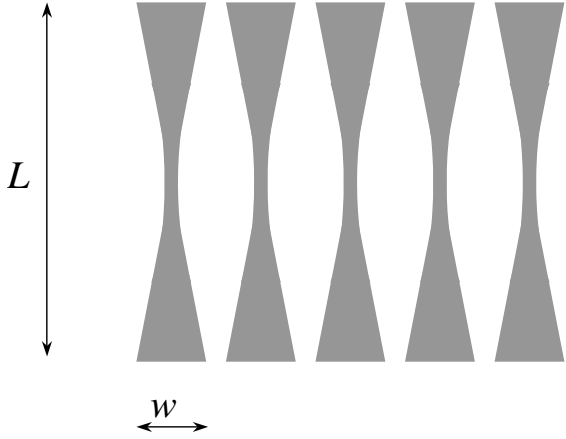,width=10cm,bbllx=70bp,bblly=250bp,bburx=450bp,bbury=500bp,clip=}
\caption{Transmitting in free space with multiple Gaussian beams}
\label{fig3}
\end{figure}

Since the bit rate cannot exceed the carrier's frequency (i.e.,
$B_{max}<\omega$) we obtain qualitatively a similar result for SC:

\begin{equation}
b_{\mathrm{free}}^{2D}=\frac{B_{max}}{w}=\frac{\omega}{2\sqrt{\lambda
L /\pi}}
\end{equation}


In the 3D case, such as an optical fiber, the SC is lower, since
the fibers should be separated in an additional dimension.

The refractive index satisfies

\begin{equation}
n(z,\mathbf{r})=\left\{\begin{array}{cc}
  n_1=n & |{\mathbf{r}}|<a \\
  n_2=1 & |{\mathbf{r}}|>a
\end{array} \right.
\end{equation}
where $a$ is now the waveguide's radius.

The propagation constant similarly satisfies \cite{Yariv}

\begin{equation}
\beta=\sqrt{q^2+n_2^2k_0^2}\simeq k_0+\frac{q^2}{2k_0}
\end{equation}

where $q^{-1}$ is the localization distance in the transversal
direction. For simplicity, let us take the most common case where
$n \simeq 1$, then \cite{Yariv}

\begin{equation}
q \simeq \frac{2}{a}\exp \left(-\frac{2}{V^2}-\gamma \right)
\end{equation}
where $V\equiv ak(n^2-1)$ is the fiber's \emph{V parameter}, and
$\gamma$ is the Euler constant. Clearly, when $ak \rightarrow 0$
we obtain

\begin{equation}
B \sim (d^2\beta/dk_0^2)^{-1/2} \sim q^{-1}
\end{equation}

However, since the area around each fiber should be proportional
to $q^{-2}$ (to avoid cross-talk in both transversal dimensions)
we finally obtain that

\begin{equation}
b=\frac{B}{Area} \sim q \sim \exp(-2/V^2) \rightarrow 0
\end{equation}

Not only does this SC converges to \emph{zero}, but it is much
lower than the SC of the free space result

\begin{equation}
b_{\mathrm{free}}^{3D} \sim \frac{\omega}{\lambda L}
\end{equation}

Note that $b_{\mathrm{free}}^{3D} L \sim \frac{\omega}{\lambda}$
depends only on the carrier frequency (or wavelength).

\bigskip

To summarize, we have calculated the spatial capacity of slab
waveguides (2D case) and fibers (3D case) in the limit of zero
width. We have shown that while the bit-rate of a single waveguide
(or fiber) is inversely proportional to its width (in the limit),
the SC decreases monotonically. In the case of a waveguide it is
shown that the SC converges to $\sim \omega/\sqrt(\lambda L)$.
This result is independent of the waveguide's refractive index,
and qualitatively similar to the SC of free 2D space.

In the case of fibers (i.e., 3D) however, the SC vanishes in the
limit of a zero width waveguide, and is therefore different from
the SC of free 3D space.

\end{document}